\documentclass[seceq]{ptptex}

\newcommand{\definition}{\stackrel{\text{d}}{\equiv}}

\notypesetlogo                       

\title{
Possible Nonlinear Completion of Massive Gravity%
}

\author{
Shinji \textsc{Hamamoto}%
}

\inst{
Department of Physics, Toyama University, Toyama 930-8555, Japan
}

\abst{
Possible nonlinear completion of massive gravity is presented. 
An additional scalar ghost contained in linear theory condensates to 
give rise to positive-energy excitations. 
}

\begin{document}

\maketitle

\section{Introduction}
In previous papers,\cite{H1,H2,H3,H4} 
we have investigated infrared regularization of 
linearized massive tensor fields. 
Two model theories have been considered: 
one is of pure-tensor (PT) type, 
which describes an ordinary massive tensor field 
of five degrees of freedom; 
the other is of additional-scalar-ghost (ASG) type, 
which contains a scalar ghost in addition to the pure tensor 
field. 
For the ASG model to have smooth massless limit, 
an auxiliary vector-like field is introduced, 
by which the original theory is promoted to the one 
invariant under vector BRS transformation. 
On the other hand, to carry out infrared regularization of 
the PT model, 
we need to introduce another auxiliary field of scalar type 
in addition to the vector-like one; 
the resulting theory is made invariant under scalar BRS 
transformation as well as the vector one. 
However, those models have to be seen as linear aproximations of 
some nonlinear theories. 
Constructing complete nonlinear theories of massive gravity 
has been left unsolved. 

The purpose of the present paper is to propose 
possible nonlinear completion of the ASG model. 
The ASG model is easier to deal with than the PT model, 
because in the former case 
only the vector BRS transformation comes on the stage. 
The nonlinear form of this transformation is nothing but the 
quantum version of general coordinate transformation. 
The scalar BRS has no classical counterpart; 
constructing its satisfactory nonlinear version requires 
further study. 

As stated above, the linear ASG model has a trouble with 
the scalar ghost. 
We will show that some nonlinear effect 
can cause background instability and ghost condensation. 
As a result, the ghost is converted to a field with 
positive-energy excitations. 

In \S 2 we summarize the results of infrared regularization 
of linearized massive tensor fields. 
In \S 3 possible nonlinear completion of massive gravity is given. 
At this stage 
the theory still has a trouble with the scalar ghost. 
Adding some nonlinear terms can settle the problem; 
the ghost condensates to leave positive-energy exitations. 
This is shown in \S 4. 
Section 5 is devoted to summary and discussions. 

\section{Linear theories}
The Lagrangian which describes a linearized massless tensor 
field $h^{\mu\nu}$ is\footnote
{The flat-spacetime metric used in the present paper is 
$\eta_{\mu\nu} = ( -1, +1, +1, +1 )$.}
\begin{equation}
L_{0} = \frac{1}{2}h^{\mu\nu}\Lambda_{\mu\nu ,\rho\sigma}h^{\rho\sigma} 
+ L_{\text{GF+FP}} ,
\label{eq:201}
\end{equation}
where $\Lambda_{\mu\nu ,\rho\sigma}$ is defined by 
\begin{multline}
\Lambda_{\mu\nu ,\rho\sigma} \definition 
\left(\eta_{\mu\rho}\eta_{\nu\sigma} - \eta_{\mu\nu}
\eta_{\rho\sigma}
\right)\square 
- \left(\eta_{\mu\rho}\partial_{\nu}
\partial_{\sigma} 
+ \eta_{\nu\sigma}\partial_{\mu}\partial_{\rho}\right)
+ \left(\eta_{\rho\sigma}\partial_{\mu}\partial_{\nu} 
+ \eta_{\mu\nu}\partial_{\rho}\partial_{\sigma}\right) . \\
\label{eq:202}
\end{multline}
A Nakanishi-Lautrup (NL) field $b_{\mu}$, a pair of 
Faddeev-Popov (FP) ghosts $(c^{\mu},\, \bar{c}_{\mu})$ 
and a gauge parameter $\alpha$ have been introduced 
into the gauge-fixing and FP-ghost Lagrangian 
\begin{align}
L_{\text{GF+FP}} & = 
b_{\mu}\left(
\partial_{\nu}h^{\mu\nu} - \frac{1}{2}\partial^{\mu}h 
+ \frac{\alpha}{2}b^{\mu}\right) 
+i\bar{c}_{\mu}\square\,c^{\mu} \nonumber \\
& = -i\delta\left[\bar{c}_{\mu}
\left(\partial_{\nu}h^{\mu\nu} - \frac{1}{2}\partial^{\mu}h 
+ \frac{\alpha}{2}b^{\mu}\right)\right] 
\label{eq:203}
\end{align}
with $h \definition h^{\mu}_{\makebox[1mm]{}\mu}$. 
The Lagrangian $L_{0}$ is invariant under the BRS transformation
\begin{equation}
\delta h^{\mu\nu} = \partial^{\mu}c^{\nu} + 
\partial^{\nu}c^{\mu} , \ \ \ \ 
\delta\bar{c}_{\mu} = ib_{\mu} .
\label{eq:204}
\end{equation}
Two-point function of the tensor fields is\footnote
{Here and hereafter spacetime coordinates are omitted 
in the field variables as well as in the $\delta$-functions.} 
\begin{multline}
\langle h^{\mu\nu}h^{\rho\sigma}\rangle = 
\frac{1}{\square}\left\{
\frac{1}{2}\left(
\eta^{\mu\rho}\eta^{\nu\sigma} 
+ \eta^{\mu\sigma}\eta^{\nu\rho} 
- \eta^{\mu\nu}\eta^{\rho\sigma}\right) \right. \\
\left. \mbox{}
- \frac{1}{2}(1-2\alpha )\frac{1}{\square}\left(
\eta^{\mu\rho}\partial^{\nu}\partial^{\sigma}
+ \eta^{\mu\sigma}\partial^{\nu}\partial^{\rho}
+ \eta^{\nu\rho}\partial^{\mu}\partial^{\sigma}
+ \eta^{\nu\sigma}\partial^{\mu}\partial^{\rho}\right)
\right\}\delta .
\label{eq:205} 
\end{multline}

Naive introduction of mass is made as 
\begin{equation}
L_{m}^{a}\left[ h^{\mu\nu}\right] = 
\frac{1}{2}h^{\mu\nu}\Lambda_{\mu\nu ,\rho\sigma}h^{\rho\sigma} 
- \frac{m^{2}}{2}\left( h^{\mu\nu}h_{\mu\nu} - ah^{2}\right) ,
\label{eq:206}
\end{equation}
where  $a$ is a real parameter taking the values 
\begin{equation}
a = 
\begin{cases}{}
\frac{1}{2} & \text{for the ASG model} , \\
1           & \text{for the PT model} . 
\end{cases}
\label{eq:207}
\end{equation}
Two-point functions are calculated as 
\begin{multline}
\langle h^{\mu\nu}h^{\rho\sigma}\rangle = 
\frac{1}{\square - m^{2}}\left\{
\frac{1}{2}\left(
\eta^{\mu\rho}\eta^{\nu\sigma} 
+ \eta^{\mu\sigma}\eta^{\nu\rho} 
- \eta^{\mu\nu}\eta^{\rho\sigma}\right) \right. \\
\left.\mbox{}
- \frac{1}{2m^{2}}\left(
\eta^{\mu\rho}\partial^{\nu}\partial^{\sigma}
+ \eta^{\mu\sigma}\partial^{\nu}\partial^{\rho}
+ \eta^{\nu\rho}\partial^{\mu}\partial^{\sigma}
+ \eta^{\nu\sigma}\partial^{\mu}\partial^{\rho}\right)
\right\}\delta
\label{eq:208}
\end{multline}
for the ASG model, and
\begin{align}
\langle h^{\mu\nu}h^{\rho\sigma}\rangle = \ & 
\frac{1}{\square - m^{2}}\left\{
\frac{1}{2}\left(
\eta^{\mu\rho}\eta^{\nu\sigma} 
+ \eta^{\mu\sigma}\eta^{\nu\rho} 
- \eta^{\mu\nu}\eta^{\rho\sigma}\right) \right. \nonumber \\
& \makebox[15mm]{}
- \frac{1}{2m^{2}}\left(
\eta^{\mu\rho}\partial^{\nu}\partial^{\sigma}
+ \eta^{\mu\sigma}\partial^{\nu}\partial^{\rho}
+ \eta^{\nu\rho}\partial^{\mu}\partial^{\sigma}
+ \eta^{\nu\sigma}\partial^{\mu}\partial^{\rho}\right) 
\nonumber \\
& \makebox[14mm]{}\left.\mbox{}
+ \frac{2}{3}\left(
\frac{1}{2}\eta^{\mu\nu} 
+ \frac{\partial^{\mu}\partial^{\nu}}{m^{2}}\right)\left(
\frac{1}{2}\eta^{\rho\sigma} 
+ \frac{\partial^{\rho}\partial^{\sigma}}{m^{2}}\right)
\right\}\delta
\label{eq:209}
\end{align}
for the PT model.
We see that the ASG model has second-order massless 
singularities, while the PT model has fourth-order.

The massless singularities of the ASG model can be eliminated 
by the following procedure: 
introduce an auxiliary vector-like field $\theta^{\mu}$, 
replace the field $h^{\mu\nu}$ with the combination 
$\left[ h^{\mu\nu} - \frac{1}{m}\left(
\partial^{\mu}\theta^{\nu} + \partial^{\nu}\theta^{\mu}\right)
\right]$
in the Lagrangian \eqref{eq:206} with $a=\frac{1}{2}$, 
and append the gauge-fixing and FP-ghost Lagrangian \eqref{eq:203}.
We then have 
\begin{align}
L_{\text{BRS}}^{a=\frac{1}{2}} & = 
L_{m}^{a=\frac{1}{2}}\left[
h^{\mu\nu} - \frac{1}{m}\left(
\partial^{\mu}\theta^{\nu} + \partial^{\nu}\theta^{\mu}\right)
\right] + L_{\text{GF+FP}} 
\nonumber\\
& = L_{m}^{a=\frac{1}{2}}\left[ h^{\mu\nu}\right]
- 2m\theta^{\mu}\left(
\partial^{\nu}h_{\mu\nu} - \frac{1}{2}\partial_{\mu}h\right) 
- \partial_{\mu}\theta_{\nu}\partial^{\mu}\theta^{\nu} 
+ L_{\text{GF+FP}} . 
\label{eq:210}
\end{align}
The BRS transformation which keeps the Lagrangian \eqref{eq:210} 
invariant is 
\begin{equation}
\delta h^{\mu\nu} = \partial^{\mu}c^{\nu} + 
\partial^{\nu}c^{\mu} , \ \ \ 
\delta\theta^{\mu} = mc^{\mu} , \ \ \ 
\delta\bar{c}_{\mu} = ib_{\mu} .
\label{eq:211}
\end{equation}
It turns out that the massless singularities in \eqref{eq:208} 
has been removed: 
\begin{multline}
\langle h^{\mu\nu}h^{\rho\sigma}\rangle = 
\frac{1}{\square - m^{2}}\left\{
\frac{1}{2}\left(
\eta^{\mu\rho}\eta^{\nu\sigma} 
+ \eta^{\mu\sigma}\eta^{\nu\rho} 
- \eta^{\mu\nu}\eta^{\rho\sigma}\right)\right. \\
\left.\mbox{}
- \frac{1}{2}\left[
(1-2\alpha )\frac{1}{\square} 
+ 2\alpha\frac{m^{2}}{\square^{2}}\right]\left(
\eta^{\mu\rho}\partial^{\nu}\partial^{\sigma}
+ \eta^{\mu\sigma}\partial^{\nu}\partial^{\rho}
+ \eta^{\nu\rho}\partial^{\mu}\partial^{\sigma}
+ \eta^{\nu\sigma}\partial^{\mu}\partial^{\rho}\right)
\right\}\delta . \\ 
\label{eq:212}
\end{multline}
The massless limit of this model can be smoothly taken to give 
\begin{equation}
L_{\text{BRS}}^{a=\frac{1}{2}}(m=0) = L_{0} - 
\partial_{\mu}\theta_{\nu}\partial^{\mu}\theta^{\nu} ,
\label{eq:213}
\end{equation}
which is invariant under the BRS transformation 
\begin{equation}
\delta h^{\mu\nu} = \partial^{\mu}c^{\nu} + 
\partial^{\nu}c^{\mu} , \ \ \ 
\delta\theta^{\mu} = 0 , \ \ \ 
\delta\bar{c}_{\mu} = ib_{\mu} .
\label{eq:214}
\end{equation}
The transformation law \eqref{eq:214} shows that 
the vector-like field $\theta^{\mu}$ behaves as four massless scalar 
fields in this limit. 
The zeroth component $\theta^{0}$ \emph{is} a ghost as seen from 
\eqref{eq:213}. 

To remove the massless singularities of the PT model, we have to 
introduce another field of scalar type $\varphi$ 
in addition to the vector-like field $\theta^{\mu}$. 
And then we replace $h^{\mu\nu}$ with 
$\left[ h^{\mu\nu} - \frac{1}{m}\left(\partial^{\mu}\theta^{\nu} + 
\partial^{\nu}\theta^{\mu}\right) 
+ \frac{2}{m^{2}}\partial^{\mu}\partial^{\nu}\varphi\right]$ 
in \eqref{eq:206} with $a=1$ : 
\begin{align}
L_{\text{BRS}}^{a=1} & = 
L_{m}^{a=1}\left[ h^{\mu\nu} - \frac{1}{m}\left(
\partial^{\mu}\theta^{\nu} + \partial^{\nu}\theta^{\mu}\right)
+ \frac{2}{m^{2}}\partial^{\mu}\partial^{\nu}\varphi\right]
+ L'_{\text{GF+FP}} \nonumber \\
& = L_{m}^{a=1}\left[ h^{\mu\nu}\right]
- 2\left( m\theta^{\mu} - \partial^{\mu}\varphi\right)
\left(\partial^{\nu}h_{\mu\nu} - \partial_{\mu}h\right) 
- \frac{1}{2}\left(\partial^{\mu}\theta^{\nu} - \partial^{\nu}
\theta^{\mu}\right)^{2} \nonumber \\
& \makebox[20em]{} + L'_{\text{GF+FP}} , 
\label{eq:215}
\end{align}
where $L'_{\text{GF+FP}}$ is the gauge-fixing and FP-ghost 
Lagrangian fixed below. 
The BRS transformation under which 
$L_{\text{BRS}}^{a=1}$ \eqref{eq:215} is invariant 
is given by 
\begin{equation}
\begin{cases}
\ \displaystyle
\delta h^{\mu\nu} = \partial^{\mu}c^{\nu} + 
\partial^{\nu}c^{\mu} , & 
\displaystyle
\delta\bar{c}_{\mu} = ib_{\mu} , \\ 
\ \displaystyle
\delta\theta^{\mu} = mc^{\mu} + \partial^{\mu}c , & 
\displaystyle
\delta\bar{c} = ib , \\ 
\ \displaystyle
\delta\varphi = mc , & 
\end{cases}
\label{eq:216}
\end{equation}
where the FP-ghost pair $(c, \bar{c})$ and the NL field $b$ 
have been newly introduced. 
They are associated with the scalar part of the transformation. 
A possible choice of the gauge-fixing and FP-ghost Lagrangian 
$L'_{\text{GF+FP}}$ is 
\begin{align}
L_{\text{GF+FP}}' & = -i\delta\left[\bar{c}_{\mu}
\left(\partial_{\nu}h^{\mu\nu} - \frac{1}{2}\partial^{\mu}h 
+ \frac{\alpha}{2}b^{\mu}\right)
+ \bar{c}\left(\partial_{\mu}\theta^{\mu} 
- \frac{m}{2}h 
+ \frac{\beta}{2}b\right)\right] 
\nonumber \\
& = L_{\text{GF+FP}}
+ b\left(\partial_{\mu}\theta^{\mu}-\frac{m}{2}h+
\frac{\beta}{2}b\right)+i\bar{c}\,\square\, c
\label{eq:217}
\end{align}
with the second gauge parameter $\beta$. 
It can be shown that the massless singularities in \eqref{eq:209} 
are removed by such a BRS extension. 
In the massless limit, this model describes a massless tensor field 
$H^{\mu\nu} \definition h^{\mu\nu}-\eta^{\mu\nu}\varphi$, 
a massless \emph{gauge} field $\theta^{\mu}$ 
and a massless scalar field $\varphi$. 
We have no ghost in this case. 

On carrying out nonlinear completion in the following sections, 
we restrict our consideration to the ASG model. 
In that case there appears only the vector BRS tranformation, 
which is just a linear version of the quantum general coodinate 
transformation. 
On the other hand, when the PT model is considered,
the scalar BRS transformation 
which does not find any counterpart in general relativity 
has to be managed as well. 
This gives rise to new troubles. 
Although the ASG model has the ghost problem, 
we can make it innocuous by ghost condensation mechanism, 
and can even make good use of it.

\section{Nonlinear completion}
To discuss nonlinear theories we introduce 
a curved-spacetime metric $g_{\mu\nu}$ 
\begin{equation}
g_{\mu\nu} \definition \eta_{\mu\nu}-\kappa h_{\mu\nu} 
\label{eq:301}
\end{equation}
with a gravitational constant $\kappa$, 
and a tetrad field $e_{k}^{\ \mu}$ associated with the metric 
\begin{equation}
e_{k}^{\ \mu}e^{k\nu}=g^{\mu\nu}. 
\label{eq:301a}
\end{equation}
The vector BRS transformation \eqref{eq:211} is immideately 
extended to its nonlinear form: 
\begin{equation}
\begin{cases}
\ \displaystyle
\delta e_{k}^{\ \mu} = \kappa\left(
\partial_{\rho}c^{\mu}\cdot e_{k}^{\ \rho}
-c^{\rho}\partial_{\rho}e_{k}^{\ \mu}\right), \\ 
\ \displaystyle
\delta\theta^{\mu} = mc^{\mu}
- \kappa c^{\rho}\partial_{\rho}\theta^{\mu}, \\
\ \displaystyle
\delta c^{\mu} = -\kappa c^{\rho}\partial_{\rho}c^{\mu}, \\
\ \displaystyle
\delta\bar{c}_{\mu} = ib_{\mu} .
\end{cases}
\label{eq:302}
\end{equation}
Basic quantities invariant under such a BRS tranformation can be 
constructed: 
\begin{gather}
E_{k}^{\ \mu} \definition 
e_{k}^{\ \mu}-
\frac{\kappa}{m}e_{k}^{\ \rho}\partial_{\rho}\theta^{\mu}, 
\label{eq:303}\\
G^{\mu\nu} \definition 
E_{k}^{\ \mu}E^{k\nu} = 
g^{\mu\nu}-\frac{\kappa}{m}\left(
g^{\rho\mu}\partial_{\rho}\theta^{\nu}
+g^{\rho\nu}\partial_{\rho}\theta^{\mu}\right)
+\left(\frac{\kappa}{m}\right)^{2}
g^{\rho\sigma}\partial_{\rho}\theta^{\mu}
\partial_{\sigma}\theta^{\nu}. 
\label{eq:303a}
\end{gather}
In fact they behave as BRS scalars: 
\begin{equation}
\delta E_{k}^{\ \mu}= 
-\kappa c^{\rho}\partial_{\rho}E_{k}^{\ \mu}, \ \ \ \ 
\delta G^{\mu\nu} = -\kappa c^{\rho}\partial_{\rho}G^{\mu\nu} .
\label{eq:304}
\end{equation}
Possible Lagrangians are therefore 
\begin{equation}
L = \sqrt{-g}\,F\left( E_{k}^{\ \mu}\right) ,
\label{eq:305}
\end{equation}
where $F$ is an arbitrary function. 
The action obtained is invariant indeed because the Lagrangian 
\eqref{eq:305} is transformed as
\begin{equation}
\delta L = -\kappa\partial_{\mu}\left(c^{\mu}L\right) .
\label{eq:306}
\end{equation}

Lagrangians which are at most quadratic in $E_{k}^{\ \mu}$ 
\emph{and} reduce to \eqref{eq:210} in the flat-spacetime limit 
$\kappa\rightarrow 0$ 
are found to consist of three terms as follows: 
\begin{equation}
L = \tilde{L}_{m}+\gamma\tilde{L}_{N}
+\tilde{L}_{\text{GF+FP}} 
\label{eq:307}
\end{equation}
with an arbitrary real number $\gamma$. 
Each of those terms is given by 
\begin{align}
\tilde{L}_{m} & = \frac{1}{2\kappa^{2}}\sqrt{-g}\left[ 
R+\frac{m^{2}}{2}\left( 2-G^{\mu\nu}\eta_{\mu\nu}\right)
\right] , 
\label{eq:308}\\
\tilde{L}_{N} & = 
\frac{m^{2}}{\kappa^{2}}\sqrt{-g}\left[
-3+2E_{k}^{\ \mu}\delta_{\mu}^{k}
-\frac{1}{2}\left(E_{k}^{\ \mu}\delta_{\mu}^{k}\right)^{2}
+\frac{1}{2}E_{k}^{\ \mu}\delta_{\mu}^{l}E_{l}^{\ \nu}\delta_{\nu}^{k}
\right] , 
\label{eq:208a}
\end{align}
and 
\begin{align}
\tilde{L}_{\text{GF+FP}} & = 
-\frac{1}{\kappa}i\delta\left[\bar{c}_{\mu}\left(
\partial_{\nu}\tilde{g}^{\mu\nu}
+\frac{\alpha}{2}\kappa\eta^{\mu\nu}b_{\nu}\right)\right]
\nonumber \\
& = 
\frac{1}{\kappa}b_{\mu}\left(\partial_{\nu}\tilde{g}^{\mu\nu}
+\frac{\alpha}{2}\kappa\eta^{\mu\nu}b_{\nu}\right) 
+ i\bar{c}_{\mu}\partial_{\nu}D^{\mu\nu}_{\ \ \,\rho}c^{\rho} ,
\label{eq:309}
\end{align}
where $\tilde{g}^{\mu\nu}$ and $D^{\mu\nu}_{\ \ \,\rho}$ 
are respectively defined by 
$\tilde{g}^{\mu\nu}\definition\sqrt{-g}\,g^{\mu\nu}$ and 
\begin{equation}
D^{\mu\nu}_{\ \ \,\rho} \definition 
\tilde{g}^{\mu\sigma}\delta_{\rho}^{\nu}\partial_{\sigma}
+\tilde{g}^{\nu\sigma}\delta_{\rho}^{\mu}\partial_{\sigma}
-\tilde{g}^{\mu\nu}\partial_{\rho}
-\left(\partial_{\rho}\tilde{g}^{\mu\nu}\right) . 
\label{eq:310}
\end{equation}
We can easily make sure that 
$\tilde{L}_{m}+\tilde{L}_{\text{GF+FP}}$ 
goes to $L_{\text{BRS}}^{a=\frac{1}{2}}$ of \eqref{eq:210} and 
$\tilde{L}_{N}$ tends to nil 
in the flat-spacetime limit $\kappa \rightarrow 0$.
In the following we take $\gamma=0$ in \eqref{eq:307}
as the simplest choice. 

Now we go further. 
Following the suggetion of Fronsdal and Heidenreich,\cite{FH} 
we identify the vaccuum expectation values of $\theta^{\mu}$ 
with the spacetime coordinates $x^{\mu}$, 
and introduce a set of Goldstone fields $A^{\mu}$: 
\begin{equation}
\theta^{\mu} = \frac{m}{\kappa}x^{\mu} - A^{\mu} .
\label{eq:311}
\end{equation}
Under the BRS transformation \eqref{eq:302}, they behave as 
four scalar fields: 
\begin{equation}
\delta A^{\mu} = -\kappa c^{\rho}\partial_{\rho}A^{\mu} .
\label{eq:312}
\end{equation}
By the use of $A^{\mu}$, $G^{\mu\nu}$ is expressed as 
\begin{equation}
G^{\mu\nu} = \left(\frac{\kappa}{m}\right)^{2}
g^{\rho\sigma}\partial_{\rho}A^{\mu}\partial_{\sigma}A^{\nu} .
\label{eq:313}
\end{equation}
It can be seen that the Lagrangian \eqref{eq:307} with $\gamma=0$ 
describes 
a spacetime with a negative cosmological constant 
$\Lambda = -\frac{m^{2}}{2}$ and four scalar fields $A^{\mu}$ 
on that background: 
\begin{equation}
\tilde{L}_{m} = \frac{1}{2\kappa^{2}}\sqrt{-g}
\left( R+m^{2}\right) - \frac{1}{4}\sqrt{-g}\,
g^{\rho\sigma}\partial_{\rho}A^{\mu}\partial_{\sigma}A^{\nu}
\eta_{\mu\nu} . 
\label{eq:314}
\end{equation}
The zeroth component $A^{0}$ \emph{is} still a ghost 
at this stage. 

\section{Ghost condensation}
The notion of ghost condensation has been introduced to obtain a 
consistent infrared modification of gravity.\cite{ACLM}
To kill the scalar ghost $A^{0}$ appearing in \eqref{eq:314}, 
we can make use of the same method. 
Let us modify the Lagrangian \eqref{eq:308} by adding further 
nonlinear terms like 
\begin{align}
\tilde{L}_{m}' & = \frac{1}{2\kappa^{2}}\sqrt{-g}\left[ 
R+\frac{m^{2}}{2}\left( 2-G^{\mu\nu}\eta_{\mu\nu}\right) + 
\frac{1}{2\kappa^{2}}Q\left( -\left(\kappa m\right)^{2}G^{00}\right)
\right] \nonumber \\
& = \frac{1}{2\kappa^{2}}\sqrt{-g}\left[ 
R+\frac{m^{2}}{2}\left( 2-G^{ij}\delta_{ij}\right) + 
\frac{1}{2\kappa^{2}}P\left( -\left(\kappa m\right)^{2}G^{00}\right)
\right] , 
\label{eq:401}
\end{align}
where 
\begin{equation}
P(X) \definition -X+Q(X) . 
\label{eq:402} 
\end{equation}
The variable $X$ denotes the negative of the kinetic term of 
$A^{0}$: 
\begin{align}
X & \definition -\left(\kappa m\right)^{2}G^{00} \nonumber \\
& = -\left(\kappa m\right)^{2}
\left[ g^{00}-2\frac{\kappa}{m}g^{0\rho}\partial_{\rho}\theta^{0}
+\left(\frac{\kappa}{m}\right)^{2}
g^{\rho\sigma}\partial_{\rho}\theta^{0}\partial_{\sigma}\theta^{0}
\right] \nonumber \\
& = -\kappa^{4}g^{\rho\sigma}\partial_{\rho}A^{0}
\partial_{\sigma}A^{0} .
\label{eq:403}
\end{align}
We then assume that $Q(X)$ starts from the second order when 
expanded into a power series of $X$ 
\begin{equation}
Q(X) = a_{2}X^{2}+a_{3}X^{3}+\cdots ,
\label{eq:404}
\end{equation}
and that $P(X)$ takes a minimum at some point $X=c^{2}$ 
\begin{equation}
\begin{cases}
\ \displaystyle
P'(c^{2})=-1+Q'(c^{2})=0,\\ 
\ \displaystyle
P''(c^{2})=Q''(c^{2})>0.
\end{cases}
\label{eq:405}
\end{equation}
The first assumption \eqref{eq:404} is for having 
flat-spacetime limit. 
In fact, if this assumption is satisfied, 
the $Q$-term of \eqref{eq:401} 
has a null effect in the limit of $\kappa\rightarrow 0$, 
and $\tilde{L}'_{m}$  of \eqref{eq:401} reaches the same limit as 
$\tilde{L}_{m}$ of \eqref{eq:308} does. 
The second assumption \eqref{eq:405} is for making the 
ghost condensate. 
Here let us expect the background spacetime to be 
of the Friedmann-Robertson-Walker type with the metric 
\begin{equation}
ds^{2}=-dt^{2}+a^{2}(t)d\Omega^{2}, 
\label{eq:405a}
\end{equation}
where $d\Omega^{2}$ is the spatial metric for a maximally 
symmetric 3-dimensional space.
Then the Goldstone field $A^{0}$ condensates once again 
under the second assumption: 
\begin{equation}
A^{0}=\frac{1}{\kappa^{2}}ct+\pi. 
\label{eq:406}
\end{equation}
Expanding $P(X)$ around the minimal point $X=c^{2}$ gives 
$\tilde{L}'_{m}$ the expression
\begin{multline}
\ \ \tilde{L}'_{m} = 
\frac{1}{2\kappa^{2}}\sqrt{-g}\left\{ R+
\left[ m^{2}-\frac{1}{2\kappa^{2}}\left( c^{2}-Q(c^{2})\right)
\right]\right\} \\
+ \frac{1}{2}\sqrt{-g}\,c^{2}Q''(c^{2})\,\dot{\pi}^{2}
+ \frac{1}{4}\sqrt{-g}\left( -g^{\rho\sigma}\partial_{\rho}
A^{i}\partial_{\sigma}A^{j}\delta_{ij}\right) + \cdots .
\ \ 
\label{eq:407}
\end{multline}
This expression shows that the scalar ghost $A^{0}$ has been 
converted to $\pi$, which has positive-energy excitations. 
We have now found that the vector-like field $\theta^{\mu}$ condensates 
to leave a quartet of the Goldstone scalars $A^{\mu}$, 
and that the zeroth component $A^{0}$ which is still a ghost further 
condensates to give rise to the field $\pi$. 
In this respect the field $\pi$ is, 
so to speak, 
a Goldstone field of the second generation. 
Note that in this case the cosmological constant 
$\Lambda= \frac{1}{4\kappa^{2}}\left( c^{2}-Q(c^{2})\right)
-\frac{m^{2}}{2}$ 
can be positive or negative 
depending on the concrete form of $Q(X)$. 

Here is the simplest example. 
Take $Q(X)$ as just a quadratic form satisfying the conditions 
\eqref{eq:405} 
\begin{equation}
Q(X)=\frac{1}{2c^{2}}X^{2}. 
\label{eq:408}
\end{equation}
Then $\tilde{L}_{m}'$ \eqref{eq:407} becomes 
\begin{multline}
\ \ \ \ \ \ \ \tilde{L}'_{m} = 
\frac{1}{2\kappa^{2}}\sqrt{-g}\left[ R+
\left( m^{2}-\frac{c^{2}}{4\kappa^{2}}\right)\right] \\
+ \frac{1}{2}\sqrt{-g}\,\dot{\pi}^{2}
+ \frac{1}{4}\sqrt{-g}\left( -g^{\rho\sigma}\partial_{\rho}
A^{i}\partial_{\sigma}A^{j}\delta_{ij}\right) + \cdots .
\ \ \ \ \ \ \ 
\label{eq:410}
\end{multline}

\section{Summary and Discussions}
In this paper we have presented a simple model of nonlinear 
completion of massive gravity. 
This model describes a spacetime with a 
positive/negative cosmological constant, 
\emph{and} a set of fields on that background. 
The set consists of a Goldstone field $\pi$ with 
positive-energy excitations and a triplet of normal massless 
scalar fields $A^{i}$. 
The pion field emerges from ghost condensation. 
Physical implications of the ghost condensation, 
which includes possible Lorentz-violating effects, 
have been discussed extensively in Ref.~\citen{ACLM}. 

This model has the smooth massless as well as the smooth 
flat-spacetime limits. 
In the flat spacetime limit, it reduces to the 
BRS-extended ASG model. 
The additional scalar ghost appearing there is just an artifact, 
telling us that the flat spacetime is a false vacuum. 

Nonlinear completion of the PT model is still unfinished. 
This problem is mainly related to how to find a nonlinear version of 
the scalar BRS transformation. 


%

\end{document}